\documentclass[aps,prd,showpacs,amssymb,floatfix,nofootinbib]{revtex4}
\usepackage{graphicx,amsmath,subfigure}

\def\be{\begin{equation}}
\def\ee{\end{equation}}
\def\bea{\begin{eqnarray}}
\def\eea{\end{eqnarray}}
\def\ltsima{$\; \buildrel < \over \sim \;$}
\def\simlt{\lower.5ex\hbox{\ltsima}}
\def\gtsima{$\; \buildrel > \over \sim \;$}
\def\simgt{\lower.5ex\hbox{\gtsima}}

\def\lesssim{\mathrel{\hbox{\rlap{\hbox{\lower4pt\hbox{$\sim$}}}\hbox{$<$}}}}
\def\gtrsim{\mathrel{\hbox{\rlap{\hbox{\lower4pt\hbox{$\sim$}}}\hbox{$>$}}}}
\def\alt{\mathrel{\hbox{\rlap{\hbox{\lower4pt\hbox{$\sim$}}}\hbox{$<$}}}}
\def\agt{\mathrel{\hbox{\rlap{\hbox{\lower4pt\hbox{$\sim$}}}\hbox{$>$}}}}

\def\gta{\ifmmode {\mathbin{\lower 3pt\hbox   
    {$\,\rlap{\raise 5pt\hbox{$\char'076$}}\mathchar"7218\,$}}}
    \else {${\mathbin{\lower 3pt\hbox
    {$\rlap{\raise 5pt\hbox{$\char'076$}}\mathchar"7218\,$}}}
    $}\fi}
\def\lta{\ifmmode {\,\mathbin{\lower 3pt\hbox   
    {$\,\rlap{\raise 5pt\hbox{$\char'074$}}\mathchar"7218\,$}}}
    \else {${\mathbin{\lower 3pt\hbox
    {$\rlap{\raise 5pt\hbox{$\char'074$}}\mathchar"7218\,$}}}
    $}\fi}

\begin{document}

\author{Stanislav Babak}
\affiliation{Albert Einstein Institute,  Am Muehlenberg 1, D-14476 Golm, Germany}
\email{stba@aei.mpg.de}
\author{Jonathan R. Gair}
\affiliation{Institute of Astronomy, Madingley Road, CB3 0HA Cambridge, UK}
\email{jgair@ast.cam.ac.uk}
\author{Edward K. Porter}
\affiliation{Albert Einstein Institute,  Am Muehlenberg 1, D-14476 Golm, Germany\footnote{Present address: APC (AstroParticules et Cosmologie), 10, rue Alice Domon et L\'{e}onie 
Duquet, 75205 Paris Cedex 13, France}}

\title{An algorithm for detection of extreme mass ratio inspirals in LISA data}

\begin{abstract}
 The gravitational wave signal from a compact object spiralling toward a massive black hole
 (MBH) is thought to be one of the most difficult sources to detect in the LISA data stream. Due to the large parameter space of possible signals and many orbital cycles spent in the sensitivity band of LISA, it has been estimated previously that  of the order of $10^{35}$ templates would be required for a fully coherent search with a template grid, which is computationally impossible. Here we describe an algorithm based on a constrained Metropolis-Hastings stochastic search which allows us to find and accurately estimate  parameters of isolated EMRI signals buried in Gaussian instrumental noise. We illustrate the effectiveness of the algorithm with results from searches of the Mock LISA Data Challenge round 1B data sets.
\end{abstract}


\maketitle


\section{Introduction}\label{sec:intro}
Extreme mass ratio inspiral (EMRI) --- a stellar mass compact object (CO) that is captured and 
spirals into a massive Black Hole (MBH) through the emission of gravitational radiation --- is 
one of the most interesting sources for the future LISA mission \cite{LISA:1998}. 
The inspiral proceeds very slowly for these sources (the inspiral rate is proportional to the mass ratio which is typically $1:10^5$) so the CO spends a significant amount of time in the strong field close to the MBH. The GW signal contains information about the geometry of the central hole which, in the case 
of a strong signal, could be extracted to confirm or otherwise that the massive objects observed in 
galactic nuclei are indeed Kerr BHs, as we suppose~\cite{ryan95,collins2004,GlamBab06,gairbumpy,AKtest}. EMRI observations may also be used to probe the stellar population in the central parsecs of galaxies, and measure the properties of astrophysical black holes to high precision~\cite{gairLISA7}. This is possible because the signal is long lived and so a coherent phase integration should recover the parameters of the binary to high accuracy, much better than anything that
will be available from electromagnetic observations. We refer  the reader to the review article \cite{AmaroSeoane:2007aw} for more details on the astrophysics that will be possible with 
EMRIs. 

In order to scope out issues associated with LISA data analysis for EMRIs, we require waveform models that are cheap and easy to generate but still capture the main features of true EMRI waveforms. One such model of the signal is the so-called analytic kludge waveform~\cite{Barack:2003fp}. It is a phenomenological template, constructed by piecing together the most important physical elements: post-Newtonian expressions for the rate of change of the orbital parameters and frequencies (Peter-Mathews approach), periastron precession, and precession of the orbital plane around the spin axis of the MBH. While these waveforms are not faithful representations, they are nonetheless representative of the true signal. This means that they should be sufficient to answer questions about what accuracy we can achieve in estimating the source parameters and at what level the confusion noise from cosmological EMRIs will be~\cite{Barack:2003fp, Barack:2006pq}. Because these waveforms are simple and fast to generate they were chosen for use in the Mock LISA Data Challenge (MLDC). The MLDC was organized to stimulate the development of data analysis tools for LISA and to establish standard notations and conventions which allow comparison of different algorithms. There have been three challenges to date~\cite{Arnaud:2006gm, Arnaud:2006gn, Arnaud:2007jy, Arnaud:2007vr, 2007arXiv0711.2667B}, which were aimed at different sources: SMBH binaries, Galactic white-dwarf binaries and EMRIs. For the two EMRI challenges five data sets were released, each containing a single EMRI signal buried in instrumental noise.
 
The five EMRI data sets had some parameters drawn from priors common to all data sets, which were: the mass of the CO $\mu \in U[9.5, 10.5]M_{\odot}$, the spin $S/M^2 \in U[0.5, 0.7]$, the plunge time 
$U[1,2]$ years and the eccentricity at plunge $e_{pl} \in U[0.15, 0.25]$; and some parameters drawn from priors that were different for each data set: MBH mass $M \in U[0.95, 1.05] \times 10^7 M_{\odot}$ (high mass binary) with SNR$\in U[40, 110]$ (1.3.1 data set),  $M \in U[4.75, 5.25] \times 10^6 M_{\odot}$ (medium mass binary) with SNR$\in U[70, 110]$ (1.3.2 data set) and with SNR$\in U[40, 60]$ (1.3.3 data set), $M \in U[0.95, 1.05] \times10^6 M_{\odot}$ (low mass binary) with $SNR \in U[70, 110]$ (1.3.4 data set) and with $SNR \in U[40, 60]$ (1.3.5 data set). Note that the last two types of signal are considered to be the most likely EMRIs to be seen by LISA: a $\sim 10 M_{\odot}$
BH falling into $\sim 10^6 M_{\odot}$ MBH in the galactic center~\cite{gairLISA7}. More details 
on the parameter sets can be found in~\cite{Arnaud:2007jy}. MLDC round 1\footnote{The ``round 1'' EMRI data set was released as part of round 2 of the MLDC.} and round 1B had the same sets of priors for the five data sets. For both challenges, it has been shown that the signal can be easily detected with high confidence, but with parameters quite different from the true ones. 

An important feature of EMRI signals is that they have many local maxima in the likelihood surface, which are quite well separated and can be as high as $75\%$ of the true maximum. Search algorithms have a tendency to find secondary maxima quickly and then get stuck there. This represents a true detection but with incorrect parameters (see \cite{2007arXiv0711.2667B, Babak:2008sn} for results). However in this article we will only regard a ``detection'' as finding the 
global maxima in the likelihood, i.e., the true source parameters. Secondary maxima are the biggest  problem for LISA data analysis. Some signals can be seen by eye in a spectrogram or in the power spectral density of the data --- the main problem is to estimate the source parameters with the best possible accuracy. A grid based search (with a sufficiently fine grid) would be guaranteed to find the 
global maxima as it covers the whole parameter space, but the required 
number of templates is so high~\cite{emrirate} that no one is presently considering doing it even with tight priors like in the MLDC. An alternative approach which has proven to be both efficient and accurate was first suggested in the context of LISA for non-spinning SMBH binary searches~\cite{Cornish:2006dt, Cornish:2006ms, Cornish:2007jv}. This approach is the semi-stochastic Metropolis-Hastings Monte-Carlo (MHMC) method, where one constructs a search chain through the parameter space (these are not in general Markovian) and follow this up by a Markov chain Monte-Carlo (MCMC) to sample the posterior distribution function. We say this approach is `semi-stochastic' since, although successive points in the chains are chosen at random, they are chosen from {\it directed} proposal distributions. This method involves generating templates as the chain moves, but its power lies in the fact that the number of points usually required to find the source is many fewer than in a full template grid. However, the chains can get stuck on local maxima. One needs to use the properties of the signal to make chains move off local maxima and explore the parameter space more widely. 

By including tricks such as simulated annealing --- ``heating'' the likelihood surface by scaling both the log-likelihood and the size of proposed jumps by a temperature factor --- and frequency annealing --- systematically increasing the range of frequencies included in the waveform template --- MHMC has solved the search problem for non-spinning SMBH binaries~\cite{Cornish:2006dt, Cornish:2006ms, Cornish:2007jv}. The full utility of MHMC for EMRI searches has so far not been demonstrated. In the round 1B MLDC release, one of the five EMRIs (of 1.3.1 type) was found by an MHMC technique~\cite{Cornish:2008zd} that used simulated annealing and began with searches on sub-segments of the data that were combined before finally running a long chain on the whole data set. The authors~\cite{Gair:2008zc} also attempted to search for EMRIs in the round 1B data using MHMC, and recovered close to the true parameters for one of the five signals (of 1.3.2 type) before the Challenge deadline. However, in both round 1 and round 1B, the best performing algorithm was not MHMC based, but a time-frequency analysis~\cite{tf1,tf1B}. Such searches are easier to implement and parameter estimation can be done if the EMRI is isolated and of sufficient brightness. However, the achievable accuracy of parameter estimation using such template-free techniques is not as good as template based methods, and the algorithm will suffer in the presence of multiple source confusion.

In this paper we describe for the first time a complete template-based EMRI search that is able to detect and recover accurate parameters for bright, isolated EMRI sources buried in instrumental noise, with parameters drawn from any of the five canonical MLDC EMRI source types. This search technique is based on our previous MHMC search, but with several improvements which we describe below.

The origin of all of the secondary maxima in the EMRI likelihood surface is in the characteristics of the signal. An EMRI signal is composed of many harmonics of the 
three fundamental orbital frequencies (of the radial $r$-motion, the azimuthal $\phi$-motion and the polar $\theta$-motion), which are evolving in time. These harmonics vary in strength (amplitude) and the local maxima arise from matching the phase of the strongest (or of several strong) harmonics for some period of time. It is possible for a signal with very different parameters to match the dominant 
harmonic very well for the whole duration of the signal but miss completely all the other harmonics.
We have tried to exploit this property by using several chains to identify the dominant harmonic and then impose a constraint between the fundamental frequencies that fixes the frequency of the dominant harmonic at some reference time.  The key idea of our search is to determine the frequency of the 
dominant harmonic/harmonics using several local maxima and this was used for the 1B submission~\cite{Gair:2008zc}. However, since the round 1B MLDC deadline, we have improved our search technique in three important ways. We have changed the parametrization of the signal, so that it is specified by the three orbital frequencies at some reference time, $t_{ref}$, and we have changed 
the proposal distribution accordingly. We use two main proposal distributions: a normal multivariate in the eigendirections of the Fisher Matrix and a variation of the Metropolis random walk which we will describe later. The second important improvement was to release the constraint after a certain point and let the chains correct the frequency of the dominant harmonic at $t_{ref}$. Finally, we have also improved the efficiency of generation of the templates
by a factor of 3--5 which has allowed us to implement an analytic maximization of the 
likelihood over the initial phases, which reduces the parameter space that must be searched. This is possible because we have developed a new type of template composed of $N$ independent harmonics
with frequency evolution defined from the analytic kludge model. From this model we can construct an $N$-dimensional $F$-statistic \cite{Jaranowski:1998qm}. This will be described later. These improvements have led to the success of the algorithm. We have analysed the ``blind'' data sets (the data sets which MLDC participants were supposed to analyze and return results for) from MLDC round 1B to tune the algorithm, and then analysed two other data sets using the search pipeline in a blind analysis. We successfully found the signal and determined the true parameters of the source for each of the seven data sets. The results of our search are summarized in 
Tables~\ref{results} and \ref{resultsBlind} to follow. In the following sections we give details of the search algorithm.

The paper is organized as follows. In section~\ref{model} we describe the signal model we have constructed for our search templates. The details of our search method are given in section~\ref{search}. We discuss the results of our search in section~\ref{resSec}, before concluding with a summary in 
section~\ref{sum}.


\section{Waveform Model}\label{model}

The analytic kludge model of EMRI signals, as used to generate the data sets for the MLDC is described
in~\cite{Barack:2003fp} and the particular implementation used for the MLDC
can be found in~\cite{Arnaud:2007jy}. For our search, we have simplified the model in order to reduce computational time. The signal can be described by harmonics of three fundamental orbital frequencies: $\nu, f_{\gamma} \equiv \dot{\tilde{\gamma}}/(2\pi), f_{\alpha} \equiv \dot{\alpha}/(2\pi)$, where a dot denotes a derivative with respect to time. The frequencies evolve according to the PN expressions
\begin{eqnarray}
\frac{d\nu}{dt} &=&
\frac{96}{10\pi}(\mu/M^3)(2\pi M\nu)^{11/3}(1-e^2)^{-9/2}
\bigl\{
\left[1+(73/24)e^2+(37/96)e^4\right](1-e^2) \nonumber \\
&&+ (2\pi M\nu)^{2/3}\left[(1273/336)-(2561/224)e^2-(3885/128)e^4
-(13147/5376)e^6 \right] \nonumber \\
&&- (2\pi M\nu)(S/M^2)\cos\lambda (1-e^2)^{-1/2}\bigl[(73/12)
+ (1211/24)e^2 \nonumber \\
&&+(3143/96)e^4 +(65/64)e^6 \bigr]
\bigr\}, \label{nudot} \\
\frac{df_{\gamma}}{dt} &=& \left[(2\pi\nu M)^{2/3} (1-e^2)^{-1}
\left[5+\frac{7}{4}(2\pi\nu M)^{2/3} (1-e^2)^{-1}(26-15e^2)\right]
-12\cos\lambda (S/M^2) (2\pi M\nu)(1-e^2)^{-3/2}\right]\frac{d\nu}{dt} \nonumber \\
&&+ \left\{6\nu(2\pi\nu M)^{2/3} (1-e^2)^{-1}
\left[1+\frac{11}{2}(2\pi\nu M)^{2/3} (1-e^2)^{-1}\right] \right.\nonumber \\&&
\left. \qquad-18\nu\cos\lambda (S/M^2) (2\pi M\nu)(1-e^2)^{-3/2}\right\}\frac{e}{(1-e^2)}\frac{de}{dt},
\label{fgamdot} \\
\frac{df_{\alpha}}{dt} &=& 2\nu (S/M^2) (2\pi M\nu)(1-e^2)^{-3/2} \left(\frac{1}{\nu} \frac{d\nu}{dt} + \frac{3e}{(1-e^2)}\frac{de}{dt}\right),\label{falphadot} \\
\frac{de}{dt}  &=& -\frac{e}{15}(\mu/M^2) (1-e^2)^{-7/2} (2\pi M\nu)^{8/3}
\bigl[(304+121e^2)(1-e^2)\bigl(1 + 12 (2\pi M\nu)^{2/3}\bigr) \, \nonumber \\
&&- \frac{1}{56}(2\pi M\nu)^{2/3}\bigl( (8)(16705) + (12)(9082)e^2 - 25211e^4 
\bigr)\bigr]\,
\nonumber \\
&&+ e (\mu/M^2)(S/M^2)\cos\lambda\,(2\pi M\nu)^{11/3}(1-e^2)^{-4}
\, \bigl[(1364/5) + (5032/15)e^2 + (263/10)e^4\bigr] ,
\label{edot}
\end{eqnarray}
The harmonic structure of the signal is best seen when using a static source frame defined by the spin of the MBH which is assumed to be constant in this model. The radiative frame is then constructed using the direction of propagation (or direction to the source from the solar system barycenter (SSB)) and the spin direction of the MBH. The advantage of those two frames is that they are static and all the time dependence is encoded in the amplitude and phases of the harmonics explicitly. In the original analytic kludge paper~\cite{Barack:2003fp}, the waveform was expressed relative to a precessing frame, tied to the orbital angular momentum, which makes it more complicated to compute the harmonic decomposition. In the static SSB frame, the signal takes the following form
\be
h \sim \mu(2\pi M \nu(t))^{2/3} \sum_{l,n,m}  A_{l,n,m}(e(t)) e^{i(n\phi(t) + l\tilde{\gamma}(t) + m\alpha(t))} 
\ee
The amplitude of each harmonic depends on the source location (ecliptic coordinates), orientation of the spin and the
orbital eccentricity. These expressions are known analytically, but are messy so we do not include them explicitly here. We have examined the amplitudes of the harmonics for a wide range of parameters and find that harmonics of the perihelion precession with $l\neq2$ are significantly suppressed. We can also neglect the contribution from harmonics of the orbital frequency with $n>5$ for orbital eccentricities less than $e \sim 0.65$. Moreover, by construction, the 
analytic kludge waveforms are quadrupolar and therefore only harmonics of the orbital plane precession frequency with $m\in[-2,2]$ are allowed. This will not be the case for real EMRI signals and more sophisticated models include higher multipoles~\cite{NK, genTB}. Knowledge of the analytic form of the harmonic amplitudes and the restriction of the number of 
harmonics, to as few as $\sim4$--8 dominant harmonics in most cases, allows us to simplify the template and make its generation more efficient. The amplitudes of the harmonics 
depend on Bessel functions, with argument, $ne(t)$, that is usually small, so a further simplification follows by expanding these as Taylor series and truncating at the desired level of accuracy. 

The technique of Time-Delay Interferometry (TDI)~\cite{lrr-2005-4} will be used in order to cancel the laser noise in the LISA data. The basics for this technique is to combine the data 
sent and received by different spacecraft with time delays chosen to cancel the common laser noise component. The LISA response function is consequently somewhat complicated, although it can be significantly simplified in the long wavelength limit $\omega_{GW} L \ll 1$ \cite{Cornish:2002rt} ($L$ is LISA's arm length $\sim 16.7$s\footnote{We are working in geometrical units $G=c=1$} and $\omega_{GW}$ is the GW frequency). For an EMRI into a lower mass MBH (1.3.4/1.3.5 type source), the frequency of the GWs can be quite high and neither the long wavelength nor rigid adiabatic approximations \cite{Cornish:2002rt} are valid. Our code uses the full response but with time delays applied only to $h_{GW}^{SSB}$ and not to the LISA motion --- treating LISA as a solid rotating triangle.
This is overkill for the high mass MBH EMRIs (1.3.1 type) (and probably for the medium mass, 1.3.2 type, sources as well), but we decided to use the same codes for all searches. We have saved on computing time for the higher mass systems by using a lower sampling rate during the integration of the orbital motion and then up-sampling while generating the TDI streams. We also use linear interpolation to compute the time delayed data from a regularly space SSB time series, rather than a more complicated and expensive interpolation scheme.

We have verified our waveform templates against full analytic kludge templates generated using {\it SyntheticLISA}~\cite{Vallisneri:2004bn}, by computing the overlap, which 
is the inner product,
\be
(s|h) = 2\int \frac{\tilde{s} \tilde{h}^* + \tilde{s}^* \tilde{h}}{S_h(f)} df,
\label{olp}
\ee
between two normalized $(s|s) = (h|h) =1$ signals. In the expression above, a tilde denotes the Fourier transform, and $S_h(f)$ is the one-sided noise 
power spectral density. We have found that the overlap between our approximate model and the accurately computed templates is in the range $[0.93-0.99]$ depending on the source parameters, in particular the mass of the MBH (the overlap is usually higher for high mass MBH EMRIs).
The loss in overlap comes primarily from mismatches in the amplitude, while the phase 
is tracked very well. This is to be expected, as we do not make any approximations in our computation of the evolution of the orbital parameters and frequencies. The small mismatch between the template and the signal will lead to a bias in the parameters estimated for the signal. A mismatch in amplitude will primarily affect the estimated signal-to-noise ratio/luminosity distance for the source, while a phase error will lead to errors in all of the intrinsic parameters. We have found that our model is very faithful, with typical model-induced parameter errors for MLDC source types being $\sim1$--$2\sigma$, where $\sigma$ is the parameter error as estimated from the Fisher-Matrix. In other words, we expect the model-error to be of similar size, but no larger than the error in parameter recovery that arises from instrumental noise in the detector. This is confirmed by the results of our search summarised in Tables~\ref{results}--\ref{resultsBlind}. We see that our parameter recovery was very good, except for the SNR which was as much as $\sim5\%$ different in two of the low mass MBH cases.

\section{Search Algorithm}\label{search}

In this section we will describe the overall search algorithm. In practice, there were some differences between the searches for each source and we will discuss these source-specific details in the next section. Our search consists of three steps. In the first step
we look for the ``footprints'' of the signal, and for points from which we can seed
our subsequent chains. In the second step we construct chains using the identified properties of the signal, via a constrained Metropolis Monte-Carlo search on the half year long segments 
of data. The final step is to narrow down the parameters of the signals by extending 
the duration of the templates to the total length of observation (this is similar to what 
was done here \cite{Cornish:2008zd}). In the following subsections we will give details
on the implementation of the three steps.

\subsection{Uniform Jumps}

As mentioned previously, the basis of our search method is to identify
as many strong local maxima in the likelihood as possible and then use the information encoded in the points to direct the search toward the true solution. The first step is very simple and remarkably efficient. In spirit it is similar to using a random template bank, as used in \cite{Messenger:2008ta}. We generate template waveforms for the last half a year of inspiral before plunge by integrating the
equations of motion backwards), and with parameters randomly chosen from within uniform
priors.
For greater efficiency we include maximization of the log likelihood over the distance, plunge time and three orbital phases at plunge. We maximize over the plunge time in the usual way, by computing the correlation of the template with the data (instead of the inner product). The maximized value of the plunge time is then used for constructing a
new filter and we compute the likelihood maximized over phases and distance.
The maximization over the distance is done in the usual way: the log-likelihood (up to a constant factor) is given by
\be
\Lambda = -\sum_{I}(x_I -h_I | x_I -h_I ) \sim \sum_I 2(x_I | h_I) - (h_I|h_I), 
\ee
where $I= \{A,E\}$ runs over orthogonal TDI streams \cite{lrr-2005-4}
which play the role here of independent detectors, $x_I = n_I + s_I$
is the corresponding TDI data which is combined out of the noise $n_I$ and a signal $s_I$;
$h_I$ is a template and the inner product is defined in Eq.~(\ref{olp}) above.

An amplitude factor, ${\cal A}$ (inversely proportional to the luminosity distance to
the source) can be factored out $h_I = A \hat{h}$ and maximized over analytically.
The maximum likelihood estimator for the amplitude is
\be
{\cal A} = \frac{\sum_I (x_I|\hat{h}_I)}{\sum_I (\hat{h}_I | \hat{h}_I)}.
\ee
and then the maximized log of likelihood is 
\be
\Lambda_{max\, ({\cal A})} = \frac{\left[\sum_I (x_I|\hat{h}_I)\right]^2}
{\sum_I (\hat{h}_I | \hat{h}_I)} \label{maxL} 
\ee
This value is sometimes referred to as SNR$^2$, since if $s_I = h_I$, it reduces to $\sum_I (h_I | h_I)$, which is the square of the matched-filtering signal-to-noise ratio. Note that Eq.~(\ref{maxL}) is not sensitive to the sign of the inner product.
 
Maximization over the three initial orbital phases is more involved. We consider a template which is constructed out of three bright harmonics only. The brightness
of each $m$-harmonic for a given set of source parameters depends only on the inclination angle $\lambda$ and on the inclination of the MBH's spin to the direction to the source from the SSB~\cite{tf1B}. The prior range on plunge eccentricity ensures that we would usually have $n=2$ and/or
$n=3$ as the dominant harmonics. This reasoning suggests we take the following harmonics
$n_0 = n_1 = 2,\;\; m_0\ne m_1,\;\;\; n_2 = 3,\;\; m_2=m_0$, with $m_0$ the brightest of $m_0$, $m_1$. The three initial phases for harmonics $h^{(i)}$ are
\be
\Phi_0^i = n_i\phi_0 + 2\tilde{\gamma}_0 + m_i\alpha_0.
\ee
Each harmonic is of the form $\cos(\Phi_0^i+ \tilde{\phi}^i(t))$ and hence may be decomposed as
\be
h^{(i)} = A^i_0\cos{\Phi_0^i} \hat{h}^{(i)}(0) - A^i_0\sin{\Phi_0^i} \hat{h}^{(i)}(\pi/2),
\ee
here $\tilde{h}^{(i)}(0)$ means taken at zero initial phase. The three-harmonic template can therefore be written 
\be
h^c = h^{(0)} + h^{(1)} + h^{(2)} = \sum_{j=0}^5 a_j h^j.
\label{3harm}
\ee
Omitting all cross harmonic terms we can analytically maximize the  likelihood of our template $h^c$ over all values of the constants $a_j$, in a similar way to the $F$-statistic,
\bea
a_{2i} = \frac{\sum_I (x_I | \hat{h}^{(i)}_I(0))}{\sum_I (\hat{h}_I^i | \hat{h}_I^i )},\;\;\;\;
a_{2i+1} = \frac{\sum_I (x_I | \hat{h}^{(i)}_I(\pi/2))}{\sum_I (\hat{h}_I^i | \hat{h}_I^i )},
\label{acoef}
\eea
where $i=0,1,2$. This leads to the following maximum likelihood estimators for the 
amplitude and phase of the harmonics:
\be
\Phi_0^j = \arctan{\left(\frac{a_{2j+1}}{a_{2j}}\right)},\;\;\;\;
A^j_0 = \sqrt{a_{2j}^2 + a_{2j+1}^2}
\label{physMax}
\ee
With the choice of harmonics given above, we can obtain the initial
orbital phases from the maximum likelihood estimates of the harmonic amplitudes and phases:
\bea
\phi_0 &=&  \frac{\Phi_0^0 - \Phi_0^2}{l_0 - l_2}\\
\tilde{\gamma}_0 &=& \frac1{2}\left[ 
-\Phi_0^0 \frac{m_0l_0 - m_1l_2}{(l_0-l_2)(m_0 -m_1)} +
\Phi^1_0 \frac{m_0}{m_0-m_1} + \Phi^2_0\frac{l_0}{l_0-l_2}
\right]\\
\alpha_0 &=& \frac{\Phi_0^0 - \Phi_0^2}{m_0 - m_1}.
\eea
After a new plunge time is determined from the correlation analysis, we estimate the initial 
phases using the above method. This guarantees that we have chosen the 
optimal phases if at least one of the harmonics in the template matches the signal. We then use the maximised phases to compute the log-likelihood, Eq.~(\ref{maxL}). For this stage of the search, the longer we run the more good points we get. We typically use about 100 - 200 CPUs for several days in this phase, and normally identify a few dozen distinct secondaries with SNR of about 20\% - 40\% of the maximum.

\subsection{Search on subsets of data}

The next stage is to split the data into half-year long segments and to run a Markov
chain Monte Carlo (MCMC) using the Metropolis rejection/acceptance rule~\cite{metrop}. The MCMC technique works as follows: given a data set $s(t)$ and a set of templates $h(t;\vec{x})$, we choose a starting point, $\vec{x}$, in the parameter space.  We then propose a jump to another point, $\vec{y}$, in the space by drawing from a certain proposal distribution, $q(\vec{y}|\vec{x})$, and evaluate the Metropolis-Hastings ratio
\begin{equation}
H = \frac{\pi(\vec{y})p(s|\vec{y})q(\vec{x}|\vec{y})}{\pi(\vec{x})p(s|\vec{x})q(\vec{y}|\vec{x})}.
\label{MHrat}
\end{equation}
Here $\pi(\vec{x})$ are the priors of the parameters, which, in our analysis, were taken to be uniform distributions within the ranges allowed by the MLDC. The function $p(s|\vec{x})$ is the likelihood
\begin{equation}\label{eqn:likelihood}
p(s|\vec{x}) = C\,e^{-\left<s-h\left(\vec{x}\right)|s-h\left(\vec{x}\right)\right>/\Theta},  
\end{equation}
where $C$ is a normalization constant and $\Theta=2$ without annealing.  This jump is then accepted with probability $\alpha = \min(1,H)$, otherwise the chain stays at $\vec{x}$. In our search, we use the Metropolis rejection/acceptance rule which simplifies the above by assuming the proposal $q(\vec{y}|\vec{x})$ is symmetric, so the ratio~(\ref{MHrat}) is just the product of the likelihood ratio with the prior ratio. In this stage we also include simulated annealing, which means that $\Theta$ is allowed to vary from $2$. This has the effect of smoothing and flattening the likelihood surface, which makes it easier for the chain to move around and climb up the surface to the maximum. The idea is to have a high heat initially, to encourage the chain to explore widely and find the global maximum, then cool the surface so the chain locks into the vicinity of the maximum. We vary the temperature as the chain advances according to a schedule of the form 
\begin{equation}
\Theta = 2 \times \left\{ \begin{array}{ll} \left({\rm SNR_0}/{\rm SNR}\right)^3
& {\rm SNR} \le {\rm SNR_0}\\
1 & {\rm SNR} > {\rm SNR_0}
\end{array}\right.,
\label{heatsch}
\end{equation}
where $SNR_0$ is typically $6-7$. This annealing scheme is used at the beginning of the search and
helps  to find the trace of the signal quickly by exploring widely in the large parameter space.
At later stages of the search we also made use of thermostated annealing, as described in \cite{Cornish:2006ms, Cornish:2007jv}, which encourages the search chains to explore the vicinity of identified maxima.

At this stage of the search, we employ a different parametrization of the template, by prescribing the three orbital frequencies at some reference time $t_{ref}$ (usually chosen in the middle of the segment) 
instead of the mass and spin of the MBH. Then we start $p$ chains, where $p$ is the number of interesting points found in step one.The second step is based on the assumption that the points found in the first stage are not far in the parameter space from maxima (local/secondary or global/primary) 
on the likelihood surface. The MCMC is efficient at finding the maxima.
We have been unsuccessful in attempts to make the chains efficiently jump between local maxima until they find the global one. Instead we will show how we can use the information stored in each maximum
to guide the search in the right direction.
 
We use two main proposals, $q(\vec{y}|\vec{x})$, in the MCMC: (i) jumps within the scaled ambiguity ellipsoid (defined by the eigenvectors and eigenvalues of the variance-covariance matrix); (ii) jumps which (almost) preserve the frequency of the dominant harmonic. Let us give some more details.
Following \cite{Balasubramanian:1995bm, Owen:1998dk} we introduce the metric on the parameter space:
\be
ds^2 = g_{\mu\nu} dx^{\mu}dx^{\nu},\;\;\;
g_{\mu\nu} = (h_{,\mu} | h_{,\nu}),
\ee
where $h_{,\mu} = \partial h/\partial x^{\mu}$ and $x^{\mu}$ are parameters of the 
template. We can determine from the metric its eigenvectors, ${\bf V}_i$, and eigenvalues, 
$\lambda_i$, and hence write $g = V^T L V, $, where $V$ is the matrix of eigenvectors and $L$ is the diagonal matrix of eigenvalues. We introduce a new parametrization: 
$$
ds^2 = dW^T dW, \;\;\;\; dW_i = dY_i \sqrt{\lambda_i}\;\;\;\;
dY = V^T dX
$$
where $dX$ is a vector of parameters. We choose a value of $ds^2$ according to a 
gaussian distribution, $|\mathcal{N}(0,1)|$, and choose the direction vector $dW$ randomly oriented on the hyper-sphere with radius $ds$. In other words this proposes jumps on the surface
of the ambiguity ellipsoid scaled according to the chosen $ds$. Another similar proposal which
was used for the search of some data sets makes normal jumps in the 
eigendirections. This latter proposal was first suggested in \cite{Cornish:2006ms}. Note that for both proposals, the jumps are further scaled by the temperature when using the simulated annealing scheme, to ensure larger jumps when the surface is ``hot''.

\begin{figure}[ht]
\includegraphics[height=0.45\textheight,keepaspectratio=true]{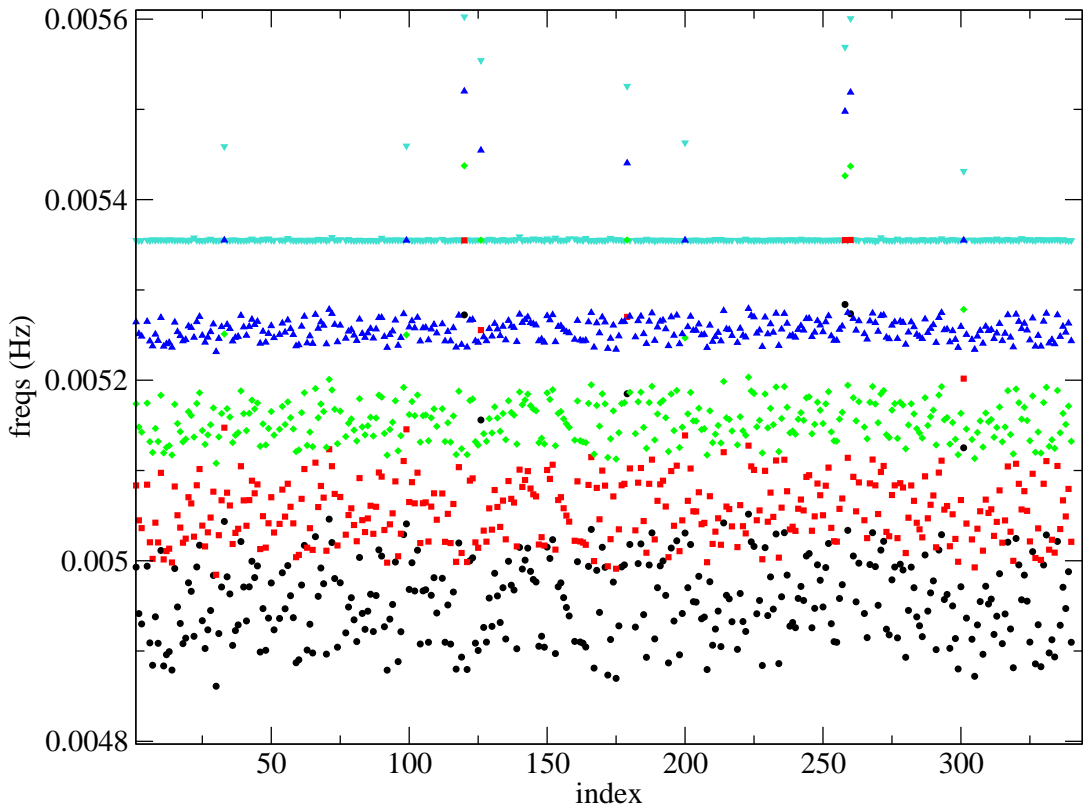}
\caption{Harmonics $l=2$, $m=[-2,2]$ for the best points identified in the first stage of the search.
The frequencies are computed at $t_{ref}$ in the middle of the third half year.
The dominant harmonic is $m=2$ (triangles down), the one with the smallest dispersion.}
\label{harms}
\end{figure}

The second proposal which we have found to be efficient at the beginning of the chain
is based on an estimation of the frequency of the dominant harmonic. For each of the high SNR points identified in the first step we can compute 
the frequency of all harmonics at the reference time, $t_{ref}$. In general, all of the points agree on the frequencies of the dominant harmonics, with a small dispersion, $\sigma$. An example from our blind search is shown in Figure~\ref{harms}.
One can clearly see that all the points agreed about the frequency of the harmonic $l=2,\;\; m=2$.
The scatter reflects the relative amplitude of the harmonics: the weakest will have the largest dispersion. One notices that some points managed to match the $m=2$ harmonic of the signal with $m=1, \;\; m=0$ or even $m=-1$ harmonic of the template (with completely wrong parameters).
The idea of this proposal is to choose $F_{l=2, m=2}$ according to a distribution $N(\overline{F_{l=2, m=2}}, \sigma_{2,2})$, where $\overline{F_{l=2, m=2}}$ and $\sigma_{2,2}$ are the mean value and the standard deviation estimated from the  points found in the first stage. Values for $f_{\alpha}$ and $f_{\gamma}$ are chosen from normal distributions in the same way. The value for $\nu$ is then defined from the 
constrain: $\nu = F_{l=2, m=2}/2 - f_{\gamma} - f_{\alpha}$. This proposal works well in 
the beginning of the chains to refine the frequencies given other parameters.

There are two reasons why we carry out the search on short duration segments at first:
(i) it is much faster to generate small templates, so we can have longer chains, (ii) the accuracy of estimating the eigenvalues and eigenvectors of the Fisher Matrix in our main proposal drops with an increase in the duration of the template.

Once the search reaches a static state, when the chains explores the posterior distribution around the local maximum,  we stop the chains. The next step is to understand what we have detected. For this purpose we change the full template to a phenomenological template which consists of $N$ independent harmonics similar to what we have used for the phase maximization (\ref{3harm}). We assume that the inner product between harmonics is zero,
 which is not really true, however it is a reasonable approximation given the 
 simplification that it allows. Each harmonic can be maximized over its amplitude and phase
 the same way as in (\ref{acoef}), (\ref{physMax}). For the physical template described 
 in section~\ref{model}, the amplitudes are functions of the spin orientation and the distance, so the amplitudes and phases are not independent. But, using this template we maximize over the amplitude and the phase for each harmonic independently (we call this a generalised $F$-statistic). This process tells us which harmonics of the template are actually detecting part of the signal. We claim a detection with a harmonic if the $SNR \ge 5$. Different chains detect different harmonics, although in the majority of cases they detect the dominant one. Usually it is easy to identify which harmonics have been detected by plotting a figure similar to Figure~\ref{harms}, although frequently the indices of the harmonic in the
template do not correspond to the indices of the harmonic of the signal that it has matched. In some cases, the chains do not agree on the identification of the harmonics and we cannot tell, for instance, whether the dominant harmonic of the signal is $m=2$ or $m=1$. In this case we run further analysis for both possibilities. Once the harmonic index of the detected harmonics have been inferred, we apply a least squares fit to determine the three fundamental orbital frequencies. Harmonics with high SNR do not always lie closer to the true frequencies than lower SNR points. We see sometimes that lower SNR points matches the frequency of the harmonics very well, but fail to fit the other waveform parameters, so the derivatives of frequency do not match.

\begin{figure}[ht]
\includegraphics[height=0.45\textheight,keepaspectratio=true]{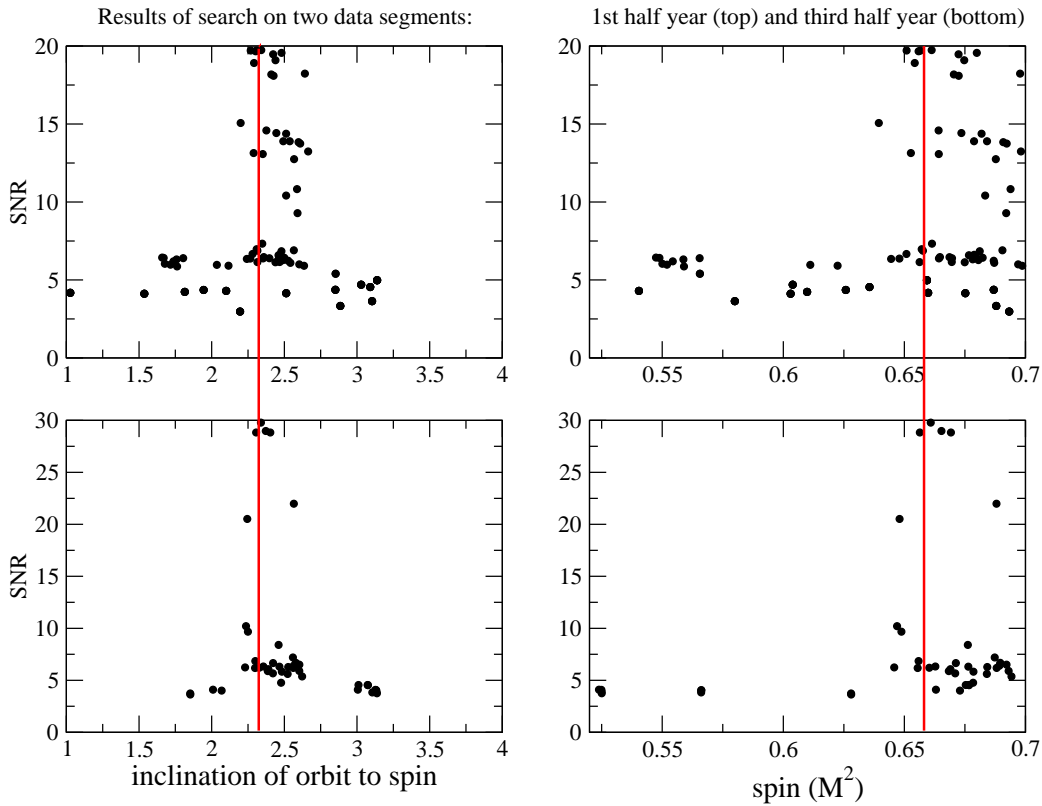}
\caption{Non-blind search for the low-mass binary. Results are shown for the search of the first (top) and third (bottom) half-year long segments. We show SNR versus parameter value for all chains for two cases: the inclination angle, $\lambda$, (left) and the MBH spin parameter (right). The vertical line in each plot is the true value of the parameter.}
\label{segmSearch}
\end{figure}

For the next step we force the orbital frequencies to be fixed at the values estimated from all the chains. We do this by turning off the jumps in the orbital frequency until the SNR has reached a value which is better than the best attained by any of the chains prior to fixing the orbital frequencies. We then release the constraint. The aim of this procedure is to first find a good guess for the frequencies, then refine the other waveform parameters (hopefully close to the true ones), then refine our estimate of the frequencies again and so on. If the initial guess is not too bad, a high SNR is achieved quite quickly when re-adjusting the other parameters. We repeat the procedure``determine frequencies -- fix them --
release'' several times if required, but this iteration need not usually be repeated more than three times.
In the figure~\ref{segmSearch} we show the final result of a run on the data from the low mass binary (1.3.4 type). In this example, the signal was found in the first and third half years of the data after only one iteration.

Before we conclude this subsection we should reiterate that we need to detect at least three
harmonics in order to be able to estimate the orbital frequencies. Moreover two of these must have different $n$-number and two must have different $m$ and the same $n$. If the harmonics detected are not sufficient  to determine the orbital frequencies, one can just use the frequency-refining proposal described above or use an $N-k$-harmonic template with the $k$ already identified harmonics excluded. This forces the search to look for other harmonics.

\subsubsection{Parameter finalization}
A good indication that we have found the signal is that the highest SNR chains cluster in parameter space. In other words, if all of the chains with SNR close (say $\gtrsim90\%$) to the maximum SNR found across all chains, have similar parameters. A further indication is that these ``best'' parameters are approximately the same for the different data sub-segments. A good example was shown in 
Figure~\ref{segmSearch} above, where we see that the first half-year and third half-year searches are producing similar results.  

\begin{figure}[ht]
\includegraphics[height=0.4\textheight,keepaspectratio=true]{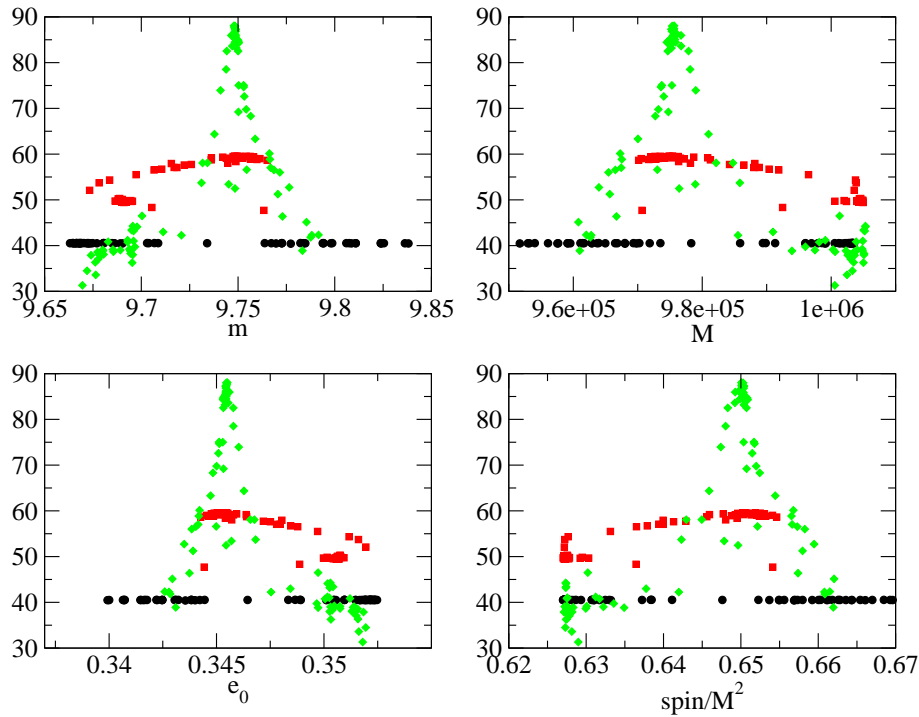}
\caption{Improvement in parameter estimation in the blind search for the low mass MBH EMRI (1.3.4 type) as we increase the duration of the template from half a year (black points) to one year (red points) to two years (green points).}
\label{improv}
\end{figure}

At this stage, the accuracy of the recovered parameters is relatively poor because we have been searching shorter data segments --- the total SNR is therefore lower and there is greater degeneracy between waveform parameters over a short duration of signal (one can accurately fit a small number of cycles in many distinct ways). The final stage of the search involves first reparameterizing the templates from all the chains in the different segments of data by their frequencies etc. at a common reference time (usually $t_{ref}=0$) and then increase the duration of the template first to one year and then to two. We then run an MCMC search with these longer waveforms, with chains starting at each of the high SNR points identified in the previous stage of the search. In Figure~\ref{improv} we show how the parameter estimation improves as we increase the template duration from one half-year to two years.

\begin{figure}[ht]
\includegraphics[height=0.4\textheight,keepaspectratio=true]{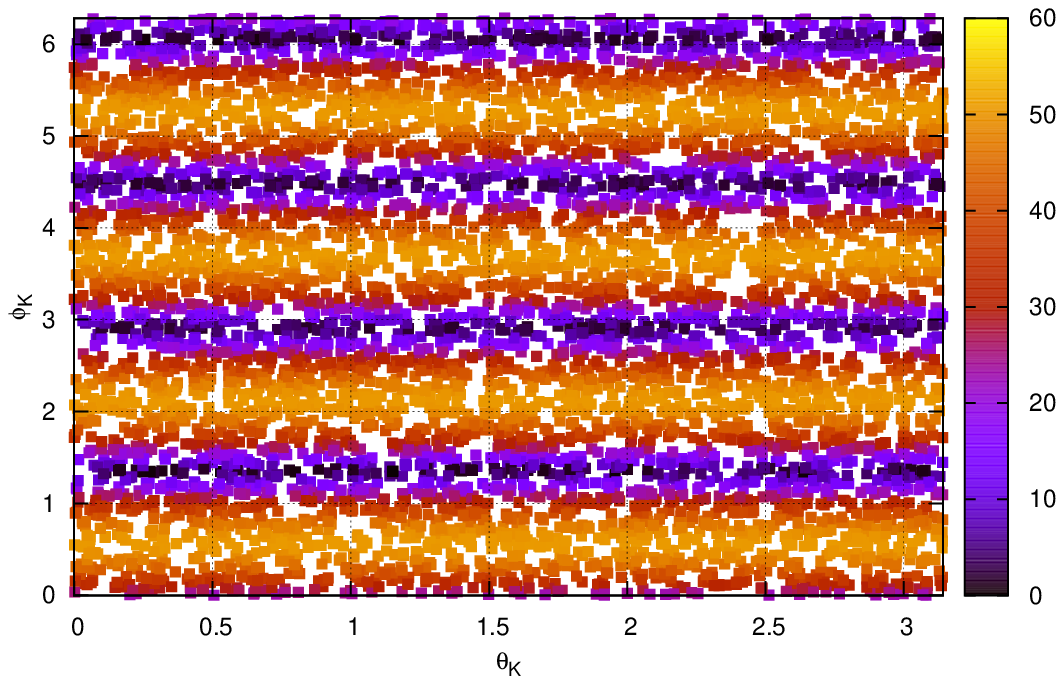}
\caption{Degeneracy in determination of the orientation of the MBH spin for the blind search of the high mass binary (1.3.1 type).}
\label{degen}
\end{figure}

In this stage, we again use the generalised F-statistic with our $N$-harmonic template to begin with. This reduces the parameter space as it does the maximization over spin orientation analytically and so it is more  efficient to use than the physical model.
There are two caveats to using this template, however. (i) If the number of harmonics included is 
large it is very slow, as we need to maximize the likelihood for each harmonic. However, 5-8
harmonics are usually enough to build up an SNR that is comparable to the full 25 harmonic SNR. (ii) The maximization leads to a smoother but larger ambiguity (error) ellipsoid in parameter space. The smoothness helps the chains to reach the global maxima, but the fact that the maximum is quite flat gives a larger error in parameter estimation. This is indicated in Figure~\ref{improv} by the wide spread in parameter values in the search with half-year long templates. To finally improve the parameter estimation we need to finish the search by using the full physical
template after we have found global maxima using the $N$-harmonic template. To seed this final analysis, we need an estimate of the spin orientation. This can be obtained from the estimated amplitudes of the harmonics, but the analytic expressions are so complicated that it would have to be done numerically. Instead, we compute the likelihood for various spin orientations and take the one giving the highest value. However, there is a four-fold degeneracy in the angle $\phi_K$, and a complete degeneracy in $\theta_K$. This is shown in Figure~\ref{degen} for the blind search for the high mass 
MBH EMRI. This plot is colour-coded by SNR, and the points were chosen randomly from a uniform distribution over the $\theta_K - \phi_K$ plane.
One can see that it is hard to distinguish between four different values of $\phi_K$
and it is almost flat in $\theta_K$. To deal with this, we ran chains for several different choices of $\theta_K,\;\; \phi_K$ and in the end took the chain with highest SNR, which did find the correct values.

\section{Results}\label{resSec}
In this section we will describe some source-specific peculiarities encountered while we were analyzing the data sets. In each case, the signal was detected at a different stage of the search algorithm described in the previous section, and we will attempt to explain why this was the case. This section 
will be divided into two subsections. The first one is dedicated to the non-blind searches which were the basis for the algorithm development and tuning. The second subsection gives details 
of the ``blind'' searches we did to test the search pipeline.
   
\subsection{Round 1B analysis}
While developing the algorithm, we analyzed the five ``challenge'' data sets that were released within the MLDC round 1B. The bulk of the search tuning and development was done after the submission deadline, so we knew parameters of the signals. We used our knowledge of the parameters only to identify when the search was going in the wrong direction so that we could try other techniques and to identify when we had detected the signal. The results of these searches, for the intrinsic source parameters, are summarized in Table~\ref{results}.

The signal was detected at different stages of the search in the various cases. The easiest signals for this algorithm to find appear to be those from medium mass MBHs (data sets 1.3.2 and 1.3.3). We believe the reason for this is that the frequency evolution of the harmonics is not so great that the harmonics are hard to detect (which is true for the low mass MBH case), but it is sufficient that the parameter space is not too degenerate (unlike the high mass MBH case which is very degenerate). The second signal in Table~\ref{results}, 1.3.2, was detected without any iterations in the "determine frequencies -- fix them -- release" phase of the search. The second medium mass binary, 1.3.3, required more work (one iteration), primarily because of the lower SNR of this source.

The high mass and low mass MBH systems are more difficult to detect. The signal from the high mass MBH EMRI has a lot of secondary maxima which are quite strong compared to the primary. These arise because the evolution in these systems is very slow, so it is easy to match harmonics for long periods of time with very different parameters. These secondary maxima are well separated and lie all over the parameter space. The analyzed signal was even worse than usual, because the inclination of the spin of the black hole to our line of sight was such that almost all of the signal power was concentrated in a single $m$-harmonic for each $n$ (we encountered a similar case in our blind search and will discuss this later).  

The low mass MBH EMRI is our canonical EMRI, as we expect these systems to dominate the event rate~\cite{emrirate,gairLISA7}. For these systems, the harmonic frequencies evolve rather significantly over the inspiral and so a template needs to match both the frequency and frequency derivative of a harmonic rather accurately in order to get high SNR. 
 This means there are fewer secondaries, but, at the same time, it also implies 
 the global maximum is rather `sharp' in parameter space, which is reflected in the
 better accuracy of recovered parameters. Usually these signals require more time on the first stage of the search (or a larger number of CPUs).  The loudest signal (1.3.4 --- fourth in the table) was more difficult to detect because of its orientation ($\lambda$ is close to $\pi/2$). This caused us to make an incorrect guess of the dominant harmonic when doing the phase maximization. We had to use an $N$-harmonic template in the search with $N=9$. The second signal (1.3.5) was not peculiar, but we had to do two iterations of the ``determine -- fix -- release'' part of the search since our first guess of the orbital frequencies was not very good due to the lower signal SNR. 

\begin{table}
\caption{\label{results}
Results of the analysis of the five ``blind'' data sets used in MLDC Challenge 1B.3. These are 1B.3.1--1B.3.5 going from top to bottom. The analysis of these data sets was not blind,  as it was mostly finished after the parameters were released.}
\begin{ruledtabular}
\begin{tabular}{c|c|c|c|c|c|c|c|c|c}
type\footnotemark & $\nu$ (mHz) & $\mu/M_{\odot}$ & $M/M_{\odot}$ & $e_0$ & $\theta_S$ &
$\phi_S$ &  $\lambda$ & $a/M^2$ & SNR\\
\hline
True & 0.1920421 &  10.296 & 9517952 & 0.21438 & 1.018 & 4.910 & 0.4394 & 0.69816  &   120.5 \\
Found &  0.1920437 & 10.288 & 9520796 & 0.21411 & 1.027 & 4.932 & 0.4384 & 0.69823 & 118.1 \\
\hline
True &  0.34227777 &  9.771 & 5215577 & 0.20791 & 1.211 & 4.6826 & 1.4358 &
0.63796 &  132.9 \\
Found & 0.34227742 & 9.769 &  5214091 & 0.20818 & 1.172 & 4.6822 & 1.4364 & 0.63804 &  132.8\\
\hline 
True & 0.3425731 & 9.697 & 5219668 &  0.19927 &  0.589 & 0.710 & 0.9282 &  0.53326 & 79.5 \\
Found & 0.3425712 & 9.694 & 5216925 & 0.19979 & 0.573 &  0.713 & 0.9298 &  0.53337 & 79.7 \\
\hline
True & 0.8514396 & 10.105 & 955795 & 0.45058 & 2.551 & 0.979 & 1.6707 & 
0.62514 &  101.6 \\
Found & 0.8514390 &  10.106 & 955544 & 0.45053 & 2.565 & 1.012 & 1.6719 &
0.62534 & 96.0\\
\hline
True & 0.8321840 & 9.790 & 1033413 & 0.42691 & 2.680 & 1.088 &  2.3196 &
0.65829 & 55.3\\
Found & 0.8321846 & 9.787 & 1034208 &  0.42701 &  2.687 & 1.053 & 2.3153 &
0.65770 & 
55.6
\end{tabular}
\end{ruledtabular}
\footnotetext{The columns are: radial orbital frequency,  mass of the CO, mass of the MBH, 
eccentricity at $t=0$, ecliptic co-latitude, ecliptic longitude, inclination angle $\lambda$,
spin of MBH, SNR recovered by template}
\end{table}

\subsection{Blind tests}

Since the high and low mass MBH EMRIs were the hardest to find, we decided to test the algorithm pipeline by performing blind tests on one data set containing  a high mass MBH EMRI, and one data set containing a low mass MBH EMRI. We use the MLDC round 1B ``training'' data sets 1.3.1 and 1.3.4 for this analysis, although we did not consult the parameter key until we had finished the search. The results  are presented in Table~\ref{resultsBlind}. We used two criteria to determine the end-point of the search and claim a detection: (i) several chains converged to the same result; (ii) the SNR of all harmonics in these best chains was comparable to and no less than the best SNR of the corresponding harmonic found in all the other chains.

For the high mass MBH signal we did not encounter any difficulties. The search with the
$N$-harmonic template resulted in quite a large error bar on the parameters because of the 
degeneracies in the parameter space and because of the relatively low SNR of this source. The degeneracy in $\theta_K,\;\; \phi_K$ for this source is illustrated in Figure~\ref{degen}.
The signal was found after one iteration of the ``determine -- fix -- release'' stage.

The signal from the low mass MBH EMRI proved much more interesting and difficult. Almost all the power was concentrated in the $m=2$ harmonic (with $n=2,3,4$).  The $F$-statistic
for 25 harmonics ($l=1,...,5$, $m=-2,...,2$) is shown for each harmonic in the matrix below

\begin{equation}
F = \left( \begin{array}{cccccc} m= & -2 & -1 & 0 & 1 & 2 \\
l=1 & 1.98 &   1.38 & 1.52 &  5.94  & 205.53\\
l=2 & 2.14 &  0.66 &  2.75  & 178.61  & 4677.62  \\
l=3 & 1.06  & 3.13 &  2.39  &  103.74  &  2109.76  \\
l=4 & 5.22  &  1.70  &  0.78  &  13.35  &  576.45 \\
l=5 & 2.85  &  1.61  &  3.27  &  4.73  &  145.50 
  \end{array}\right).
\end{equation}
We were only able to detect three $m=2$ harmonics ($n=2,3,4$) with the chains, but a second $m$-harmonic is required in order to estimate the three frequencies. To achieve this, we used the method mentioned above: we constructed an $N$-harmonic template that did not include the harmonics which had already been identified in the search. This allowed us to find the $n=2, \;\; m=1$ harmonic and hence make a preliminary estimation of all three orbital frequencies. We then needed three iterations of the ``determine -- fix -- release'' search to reach the final answer. At this stage, we were confident about the quality of the detection and, when we compared to the true parameters, we had indeed reached a very high accuracy for all parameters.

\begin{table}
\caption{\label{resultsBlind}
Results of the blind analysis of two data sets. For this analysis we used the MLDC 1B.3.1 and 1B.3.4 ``training'' data sets. Our analysis was blind in the sense that the search was run end-to-end without reference to the parameters used to generate the data sets. Our results were then compared to the known parameters at the end.}
\begin{ruledtabular}
\begin{tabular}{c|c|c|c|c|c|c|c|c|c}
type & $\nu$ (mHz) & $\mu/M_{\odot}$ & $M/M_{\odot}$ & $e_0$ & $\theta_S$ &
$\phi_S$ &  $\lambda$ & $a/M^2$ & SNR\\
\hline
True & 0.1674472 & 10.131 & 10397935 & 0.25240 & 2.985 & 4.894 & 1.2056 &
0.65101 & 52.0 \\
Found & 0.1674462 & 10.111 & 10375301 & 0.25419 & 3.023 & 4.857 & 1.2097 & 
0.65148 & 51.7 \\
\hline
True & 0.9997627 & 9.7478 &  975650 & 0.360970 & 1.453 & 4.95326 & 0.5110 & 
0.65005 &   122.9\\
Found & 0.9997626 & 9.7479 & 975610 & 0.360966 & 1.422 & 4.95339 &  0.5113 &
0.65007 & 116.0
\end{tabular}
\end{ruledtabular}
\end{table}

\section{Discussion}\label{sum}
We have described an algorithm for the detection of EMRI signals in LISA data. This algorithm is based on running multiple Markov Chain Monte Carlo search chains simultaneously. However, it relies on the key refinement that the properties of the secondary solutions that all of the chains have identified are used together to constrain further movement of the chains. It is clear from the results in Tables~\ref{results}--\ref{resultsBlind} that the algorithm is able to robustly and accurately find the true solution, in the simplified situation that we are searching for a single, bright EMRI source buried in gaussian instrumental noise. We have successfully found the source in seven out of seven mock data streams, even when the parameters were such that the waveform had unusual features, e.g., orbital inclination close to $\lambda=\pi/2$. This gives us reason to expect that the algorithm will work equally well in all comparable situations, independent of the parameters of the source.

This is the first algorithm to be published in the literature that has been demonstrated to be able to detect and determine parameters for a ``typical'' EMRI signal --- a $10M_{\odot}$ black hole falling into a $10^6M_{\odot}$ black hole --- which we expect to dominate the LISA event rate~\cite{emrirate,gairLISA7}. It is particularly gratifying that our parameter recovery is now reaching the theoretical level that was estimated from Fisher Matrix analyses~\cite{Barack:2003fp} -- MBH mass and spin determinations at the level of $10^{-4}$, and sky position accuracies of $10^{-3}$. While we would expect the Fisher Matrix to accurately represent the shape of the global maximum for these high SNR sources, it is a purely local analysis and therefore does not account for the presence of secondary maxima. The fact that our algorithm can now find the global maxima from among the bright secondaries bodes well for using EMRI sources for high precision astrophysical observations~\cite{AmaroSeoane:2007aw,gairLISA7}.

The algorithm can still be improved further. The main problem at present is that the identification of secondaries in order to determine constraints on the orbital frequencies is done by hand. We stop the search chains after stage 1, and then by hand examine the results in order to estimate suitable proposal distributions for the frequencies in the second stage. In principle this could all be done automatically, although it would require communication between different chains. If each chain had information about the best points found by all the other chains, then an adaptive proposal could be constructed from this information. The resulting search would be more like a population MCMC search. The other area of the search that could potentially be improved is the way in which we use annealing. The annealing scheme was borrowed almost verbatim from SMBH searches~\cite{Cornish:2006dt} and we have not attempted to optimize it for the EMRI problem. While this is clearly not affecting the ultimate convergence of our search, the convergence speed might be improved by modifying the annealing scheme. Other MCMC variants, for instance parallel tempering, might also improve the efficiency of the search. Parallel tempering has been demonstrated in LISA searches to great effect~\cite{KeyCornish}, but we have no immediate plans to include it in the search pipeline.

The algorithm as described here has only been demonstrated for a significantly simplified scenario --- detection of a single, bright EMRI buried in purely instrumental gaussian noise. The real LISA data stream will be very different, and is expected to contain many thousands of resolvable signals which will be overlapping in time and frequency, in addition to a noise foreground from galactic compact binaries and non-gaussian instrumental artefacts etc. It is not clear how well this search will perform under those circumstances, since it relies on being able to identify all the secondary peaks in the likelihood surface that are associated with the same signal. The relative SNRs and track shapes will provide powerful discriminators for this purpose, but there will inevitably be problems distinguishing a dim sideband harmonic of a bright source from the dominant harmonic of a similar but much more distant source. The best way to explore these complications is to attempt to analyse more realistic data sets. The next round of the MLDC, Challenge 3, includes an EMRI data set that contains five overlapping signals of low SNR. We will begin to explore source confusion by using that data set as a test case. As future MLDC releases become increasingly realistic, we will analyse them in order to demarcate where this algorithm fails and how it can be improved to cope with this greater realism.

\begin{acknowledgments}
The work of SB and EKP was supported in part by DFG grant SFB/TR~7
``Gravitational Wave Astronomy'' and by DLR (Deutsches Zentrum f\"ur
Luft- und Raumfahrt).  JG acknowledges support from the Royal Society and thanks the Albert Einstein Institute for hospitality and support while this work was being completed. 
\end{acknowledgments}

\bibliographystyle{apsrev}
\bibliography{EMRI1BSearch}



\end{document}